\documentclass[11pt,twoside,epsf]{article}
\usepackage[dvips]{graphicx}
\usepackage{rotating}
\begin{document}

{\huge \bf SS-433/W50 at TeV Energies}

\vspace{1.cm}
{ \Large Gavin P. Rowell$^*$ for the HEGRA Collaboration}
\vspace{1. cm}                                                                   

\begin{center}
{\it \Large $^*$Max Planck Institut f\"ur Kernphysik \\
                  D-69029 Heidelberg, Germany}
\end{center}

\begin{abstract}
 The HEGRA CT-System was used to search for TeV gamma-rays from the microquasar and supernova remnant
 combination SS-433/W50
 over the 1998 and 1999 observing seasons, resulting in about 20 hours of useful data. An analysis
 of various extended sources of location guided by X-Ray observations reveals no 
 evidence for TeV emission at photon energies above 1 TeV, with resulting 99\% upper limits in the range
 8 to 10\% of the Crab flux. For one of these locations, the
 eastern jet termination, a model predicting a presently detectable TeV gamma-ray flux
 via the inverse Compton upscattering of microwave background photons is now somewhat constrained.
 The results and model contraints are briefly discussed.
\end{abstract}

\section{Introduction}
The remarkable object SS-433 is a galactic microquasar \cite{Mirabel:1} powered by a compact object driving a parsec-scale jet.
The jet is possibly associated with the supernova remnant W50. A strong motivation for observations at TeV $\gamma$-ray energies exist 
by virture of the extraordinary kinetic
{\em quiescant} power output of the jets in the range $10^{39}$ to $10^{41}$ erg $s^{-1}$ \cite{Margon:1,Kawai:1,Safi-Harb:1}. This power 
amounts to a significant fraction of that required to maintain the galactic cosmic-ray (GCR) flux at Earth 
(few$\times 10^{41}$ erg s$^{-1}$ \cite{Blandford:1}). 
SS-433/W50 may therefore be one of the most powerful, and indeed significant sites of GCR acceleration in our Galaxy.

\begin{figure}
 \vspace{10cm}
 \includegraphics{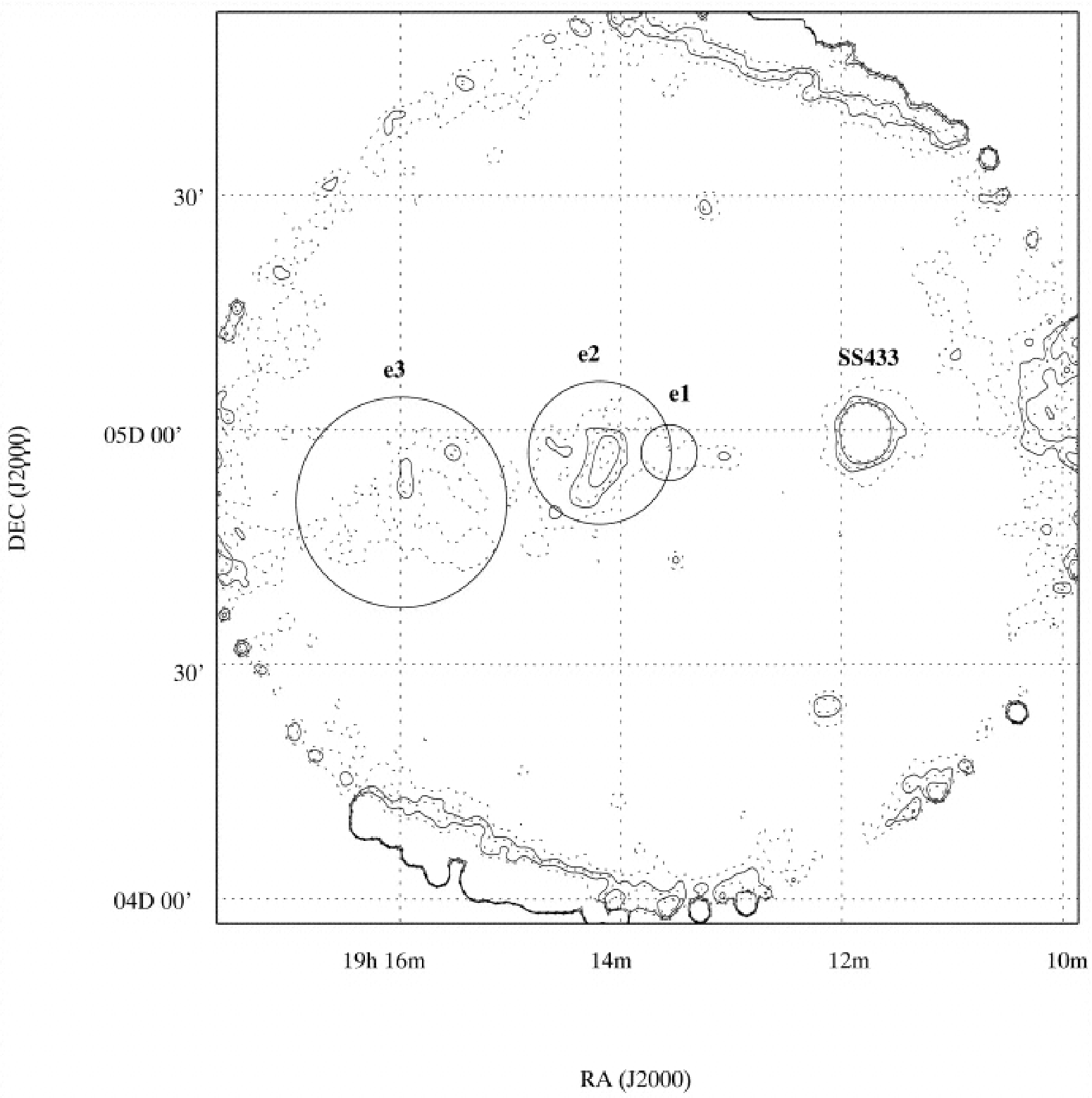}
 \includegraphics{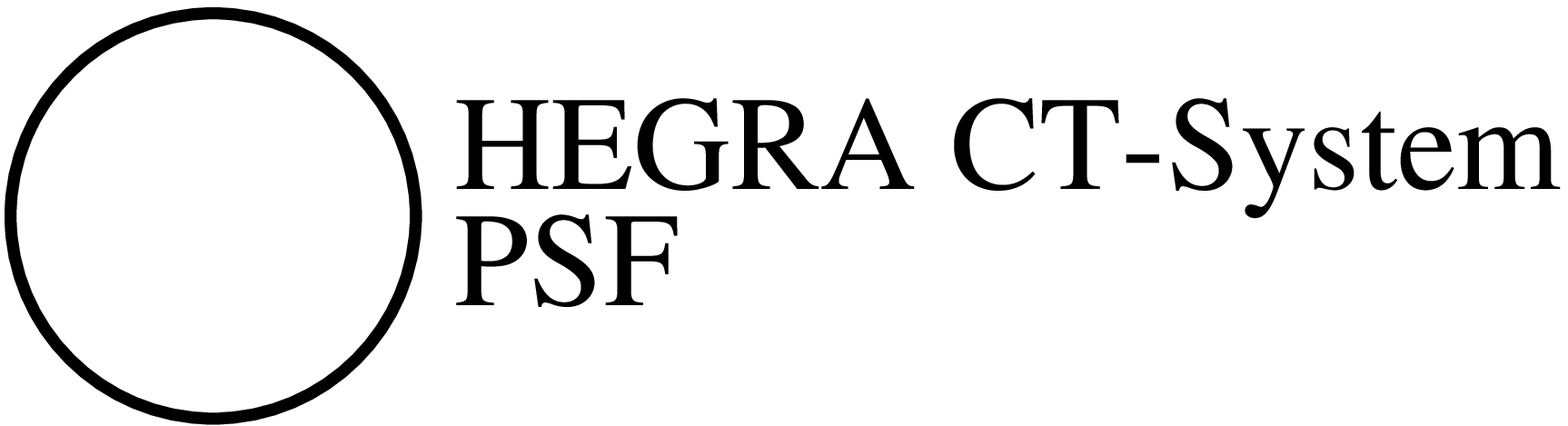}        
 \caption{ROSAT PSPC image (0.1 to 2.4 keV) revealing the eastern side of the SS-433/W50 association. Defined are a number of 
          areas (e1, e2, e3)  where separate spectral X-ray analyses were performed \cite{Safi-Harb:1}. Included is a circle depicting the
          PSF (with radius, $\sigma$=0.1$^\circ$) for a point source of TeV $\gamma$-rays as imaged by the HEGRA CT-System. The tracking
          position for HEGRA data discussed in this work is the centre of the e2 circle.}
 \label{fig:ROSAT}         
\end{figure}

Aharonian \& Atoyan \cite{Aharonian:1} (hereafter, AA) have suggested that an appreciable flux of TeV $\gamma$-rays may be produced
by inverse Compton scattering of cosmic microwave background (CMB) photons at the eastern jet termination shock, $\sim 60^\prime$ 
east of SS-433 (labelled as 'e3' \cite{Safi-Harb:1}. See figure~\ref{fig:ROSAT}), representing an interaction between the eastern jet
particles and interstellar medium (ISM). A non-relativistic shock of the type found in SNR may be capable of accelerating ISM  
particles to multi-TeV energies, and is suggested by the spatial correlation between radio and very steep 
($\Gamma=3.7^{+2.3}_{-10.7}$) X-ray synchrotron emission (from ROSAT \& ASCA data) \cite{Safi-Harb:1,Brinkmann:1,Dubner:1}. 
The X-ray emission from e3 however may also be of a thermal origin (although it is featureless!) since a power-law and thermal 
Bremsstrahlung relation are equally well fit, when considering data only from ROSAT and ASCA at energies below 10 keV.
Strong motivation for concentrating our effort on the jet
termination region is the fact that most of the jet energy is mechanical in nature \cite{Kawai:1}, leading to significant power
deposit at such regions. For example taking the estimated energy in relativistic electrons accounting for the e3 X-ray flux 
at $\leq400\times10^{46}$ erg \cite{Safi-Harb:1}, the total power in relativistic particles (electrons + baryons) at e3 amounts to
$\leq10^{39}$ erg s$^{-1}$, for a source lifetime of 10$^4$ yrs. And finally, a detection of TeV $gamma$-ray emission linked to e3,
would unambiguously prove that the shock is capable of accelerating electrons to multi-TeV energies, since alternative hypotheses
to the synchrotron model for hard X-ray tails do exist (e.g. thermal bremsstrahlung and see also \cite{Laming:1}).

Moving closer to SS-433 along the eastern jet, a hardening of
the X-ray spectrum is noticed. The power law fit to the inner regions, labeled e1 \& e2 \cite{Safi-Harb:1} is favoured only
after including X-ray data from RXTE at energies greater than 10 keV \cite{Safi-Harb:2}.
X-ray emission at e1 and e2 may result from direct synchrotron \cite{Safi-Harb:1}, inverse Compton, and/or  
synchrotron-self-Compton emission. The latter two options considered by \cite{Band:1} relating to flaring episodes were found to require 
a $B$-field well below equipartition values (in this scenario $\sim$0.7 G). 
The direct synchrotron option would also predict a spatial correlation at radio energies, {\em not presently seen}.
These inner regions are therefore more complicated to interpret than that of e3. 

\section{Observations \& Results}
Data were collected using the HEGRA CT-System\footnote{High Energy Gamma Ray Astrophysics Cerenkov Telescope}, a series of 5 atmospheric 
Cerenkov telescopes operating in coincidence. Each telescope uses a multi-pixel camera of photomulitplier tubes (PMTs) to 
parametrise the angular distribution of extensive air shower (EAS) Cerenkov radiation initiated by $\gamma$-ray primaries. 
The CT-System is situated on the Roque de los Muchachos at La Palma (2200m asl, $28^\circ \, 45^\prime$N 
$17^\circ \, 54^\prime$W).
The coincident trigger ensures a minimum of 2 telescopes view the EAS, and hence accurate reconstruction of the arrival 
direction on an event-by-event basis \cite{Konopelko:1}. At epochs covering SS-433 observations described here, the CT-System 
achieved an angular resolution, the standard deviation of a Gaussian fit (point spread function, PSF) of 
$\sim$0.1$^\circ$ at a modal energy threshold of 1.0 TeV for $\gamma$-rays. The detection of $\gamma$-ray images must be
made against a strong background of cosmic-ray-initiated EAS. This background may be suppressed by exploiting the physical differences
in EAS development between $\gamma$-ray and CR EAS, leading to differences in image shape. CR EAS are also isotropic in terms
of arrival direction and therefore will be distributed as an essentially flat background against a source of $\gamma$-rays.
The so-called {\em mean-scaled-width} ({\em msw}) parameter is used to reject CR images according to shape. A parameter, $\theta$,
the angular difference between the source and reconstructed arrival position provides CR rejection based on direction.
Details of the direction reconstruction 
may be found in \cite{Hofmann:1} and overall rejection efficiencies in \cite{Konopelko:1}. Optimal values for $\theta$ and {\em msw}, 
selected from results of Monte Carlo simulations, have been validated against the steady TeV emission of the Crab, and variable emission
from Mrk 501 \& Mrk 421. 

A total of 19.3 hours observations were accepted for analysis following removal of data adversely affected by weather conditions.
Not included in this report are results from a more recently completed scan of the Galactic Plane which included SS-433 
\cite{Aharonian:4}. Overall results will be discussed in a future paper. 
Observations were carried out using a series of ON source and OFF source tracking modes. OFF source data were taken from regions
separated in right ascension such that matching elevation bands\footnote{The energy threshold of the HEGRA CT-System is dependent on
the elevation} were achieved in both ON and OFF source data. A recently developed method \cite{Rowell:1} that 
enables the background to be determined from ON source observations is actually used in these analyses. This method estimates the
background from larger statistics than that of OFF source observations (by roughly a factor of 3), and is temporally and spatially 
consistent with ON source data.
A small systematic bias introduced by this method amounting to a $<\pm10 $\% over/under-estimation of the background is easily 
parametrised and therefore corrected.
The field of view (FOV) of the HEGRA CT-System with a radius $\sim 1.5^\circ$ (trigger efficiency $\geq$80\% of the on-axis value), 
adequately covers the eastern jet region imaged
by ROSAT. For the analysis of HEGRA data, a number of regions in the FOV were chosen {\em a-priori} as extended sources of a size 
guided by results at X-ray energies, namely, the regions e1 to e3, and SS-433 itself. For an extended source of radius $\theta_s$,
the optimal cut on event direction used in this analysis, $\theta_{op}$, is given by $\theta^2_{op} = \theta^2_{pt} + \theta^2_s$. 
$\theta_{pt}$ is the optimal cut for a point source and is a factor $\sim$1.7 larger than the PSF radius, under background-dominated 
statistics. Table~\ref{tab:results} presents the results of these analyses for each location within the FOV.
\begin{table}
 \begin{center}
 \begin{tabular}{lcrrcrr} \hline \hline
  Source & $\theta_{op}$ (deg) & ON & OFF & ON$-$OFF$^a$ ($\sigma$) & $^b \phi^{99\%}_{crab}$ & $^c \phi^{99\%}_{ph}$   \\ \hline 
  e1     & 0.13 & 155 &  420 & +2.8   & \bf 0.12 & \bf 0.20 \\
  e2     & 0.21 & 342 & 1125 & +1.2   & \bf 0.12 & \bf 0.22 \\
  e3     & 0.28 & 556 & 1953 & +0.1   & \bf 0.12 & \bf 0.21 \\
  ss-433 & 0.15 & 166 &  541 & +0.7   & \bf 0.08 & \bf 0.14 \\ \hline \hline
  \multicolumn{7}{l}{\tiny a. Statistical excess from eq. 9 of \cite{Li:1}. Norm. factor $\alpha$ = 0.283}\\
  \multicolumn{7}{l}{\tiny b. $\phi^{99\%}_{crab} = $ 99\% upper limit in units of Crab flux above 1 TeV, 1.75$\times 10^{-11}$ 
                                                                                      ph cm$^{-2}$s$^{-1}$ \cite{Aharonian:2}}\\
  \multicolumn{7}{l}{\tiny c. $\phi^{99\%}_{ph} = $ 99\% upper limit ($\times 10^{-11}$ ph cm$^{-2}$s$^{-1}$)}\\
 \end{tabular}
 \end{center}
 \caption{Results of a search for $\gamma$-ray emission above 1 TeV at various source locations around the eastern lobe of 
          SS-433/W50. See text for descriptions of source names.}
 \label{tab:results}
\end{table}

\section{Discussion}
\begin{figure}.
  \vspace{7cm}
  \includegraphics{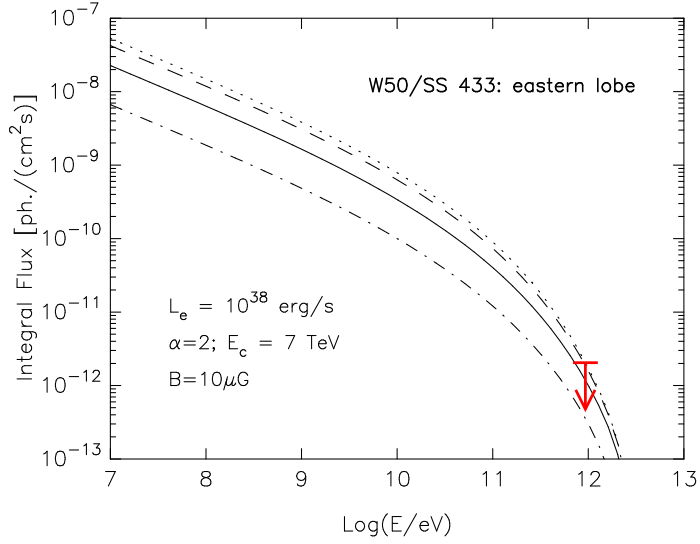}
  \caption{Integral flux of inverse Compton emission expected from the SS-433/W50 eastern lobe e3 region for different source radii
           \cite{Aharonian:1}:
           0.1$^\circ$ (dot-dashed), 0.25$^\circ$ (solid, actual radio 'ear' size), 0.5$^\circ$ (dashed) and 2.0$^\circ$ (dotted).
           Our upper limit is included and assumes a source radius of 0.25$^\circ$.}
  \label{fig:ULcomp}
 \end{figure}

 AA have argued that
 the inverse Compton scattering of Cosmic Microwave Background (CMB) photons by the same multi-TeV electrons that produce the X-ray
 synchrotron flux could lead to an appreciable TeV $\gamma$-ray emission in the range 10$^{-12}$ to 10$^{-11}$ erg cm$^{-2}$ s$^{-1}$.
 Such levels are within the sensitivity range of current ground-based instruments for reasonable observation time (10 to 40 hours).
 The radio spectral index, $\alpha \sim 0.5$ \cite{Downes:1} (a value of $\alpha \sim 0.68$ is calculated by \cite{Dubner:2}, see also
 \cite{Dubner:1}) and 
 spatial correlation of the radio and X-ray fluxes support the notion 
 that diffusive shock acceleration of electrons to TeV energies is taking place at e3. AA find that an electron injection
 spectrum of the form $N(E) \sim E^{-2} \exp(-E/7 \,{\rm TeV})$ fits well the radio and X-ray fluxes.
 Determination of the electron injection rate is tied to the $B$-field (in this model, considered a free parameter) since the radio 
 flux $\propto B^{1.5} L_e$ where $L_e$ (erg s$^{-1}$) is the injection power of electrons. AA also consider a number of source sizes,
 concluding that the measured size of the radio lobe corresponding to e3, of radius $\sim 0.25^\circ$, is consistent with a diffusion
 coefficient implied from cosmic-ray transport in the galactic disk.
 A comparison of our upper limit for e3 with the predictions of AA (for various source sizes) is presented in 
 figure~\ref{fig:ULcomp}. 

 Our upper limit is close to the prediction of AA assuming a source radius of 0.25$^\circ$, $B$-field of 10 $\mu$G, and injection rate
 of $L_e$=10$^{38}$ erg s$^{-1}$. Using the direct relationship between the expected IC ($f_{\gamma}$) and X-ray energy fluxes 
 ($f_x$) arising from the {\em same} electrons:
 \begin{equation}
  \frac{f_x(\geq \epsilon \,{\rm keV})}{f_\gamma (\geq E \,{\rm TeV})} \sim 10 \left(\frac{B}{10^{-5}\, \rm G}\right)^2
  \label{eq:ICandXray}
 \end{equation}  
 we can establish a condition on the magnetic field $B$ of the SNR. The mathematical caveat here is that the correct energy range in 
 the TeV and keV regimes must be adhered to and that we assume that the emission regions of both components have the same size as in the
 model of AA. The IC 
 ($E$ TeV) and X-ray ($\epsilon$ keV) synchrotron energies in Eq.~\ref{eq:ICandXray} are coupled according to  \cite{Aharonian:4}:
 \begin{equation}
 \epsilon \sim 0.07 \left( \frac{E}{1\, \rm TeV} \right) \left( \frac{B}{10^{-5}\, \rm G}\right) \, \rm keV
 \end{equation}
 so that for fluxes at $E \sim$1 TeV, a comparison with the X-ray flux at $\epsilon \sim$0.1 keV is required for reasonable values of 
 $B \sim 10$ to 100 $\mu$G expected after shock compression. Combining the X-ray flux at 0.1 keV with our TeV upper limit in 
 Eq.~\ref{eq:ICandXray} leads to a {\em lower limit} on the magnetic field within e3 of $B \geq 13$ $\mu$G. Using an X-ray flux 
 at slightly higher energies than 0.1 keV would naturally reduce this lower limit to values nearing that expected of the interstellar
 medium. The equipartition field for e3 is estimated at $B \sim$ 20 to 60 $\mu$G \cite{Safi-Harb:1} (using a source region of radius 
 $\sim 25$ pc, or $\sim$0.25$^\circ$ at 5.5 kpc), and so clearly, we require a 
 tighter upper limit in the TeV regime, in order to make more definite conclusions about the $B$-field in this region.

\section{Conclusion}
 A search for TeV $\gamma$-ray emission from the SS-433/W50 region, focussing on the termination-shock region 'e3', has revealed a
 99\% upper limit of 0.12 Crab flux units. The lower limit on the $B$-field at e3 determined within the framework 
 of a synchrotron and inverse-Compton model \cite{Aharonian:1} is not high enough to contrain further the range of equipartition values for
 $B$ as suggested by the X-ray results from ASCA and
 ROSAT \cite{Safi-Harb:1}. Further observations on SS-433
 with the HEGRA CT-System, recently completed, will be included in future work in order to improve our $B$ constraint.
 Finally, with their predicted sensitivities at 1 TeV, the next generation ground-based instruments HESS, VERITAS and MAGIC 
 \cite{Hofmann:2,Lorenz:1,Krennrich:1} will almost certainly resolve the issue of whether or not the e3 region is a site of electron 
 acceleration to multi-TeV energies.

\end{document}